\documentclass[traditabstract]{aa}

\usepackage{amsmath}
\usepackage{natbib}
\usepackage{graphicx}
\usepackage{txfonts}
\usepackage[]{hyperref}
\usepackage{color}

\newcommand{\src}{XTE~J1946+274}
\newcommand{\rxte}{\textit{RXTE}}
\newcommand{\Ecut}{\ensuremath{E_{\mathrm{cut}}}}
\newcommand{\Efold}{\ensuremath{E_{\mathrm{fold}}}}
\newcommand{\chisq}{\ensuremath{\chi^2_{\mathrm{red}}}}

\newcommand{\LxE}{\ensuremath{\mathrm{photons}\,\mathrm{s}^{-1}\,\mathrm{cm}^{-2}}}

\title{The reawakening of the sleeping X-ray pulsar \src}

\author{
 Sebastian M\"uller \inst{1}
 \and Matthias K\"uhnel \inst{1}
 \and Isabel Caballero \inst{2}
 \and Katja Pottschmidt \inst{3}
 \and Felix F\"urst \inst{1}
 \and Ingo Kreykenbohm \inst{1}
 \and Macarena Sagredo \inst{1,4}
 \and Maria Obst \inst{1}
 \and J\"orn Wilms \inst{1}
 \and Carlo Ferrigno \inst{5}
 \and Richard E. Rothschild \inst{6}
 \and R\"udiger Staubert \inst{7}
}

\institute{
      Dr. Karl Remeis-Observatory \& ECAP, Universit\"at 
      Erlangen N\"urnberg, Sternwartstr.~7, 96049 Bamberg, Germany
 \and CEA Saclay, DSM/IRFU/SAp-UMR AIM (7158) CNRS/CEA/Universit\'e
      Paris 7, Diderot, 91191 Gif sur Yvette, France 
 \and CRESST and NASA Goddard Space Flight Center, Greenbelt, MD 20771, USA and
      Center for Space Science and Technology, UMBC, Baltimore, MD 21250,
      USA 
 \and Department of Physics, Florida International University, Miami,
      FL 33199, USA
 \and ISDC Data Center for Astrophysics, University of Geneva, 16
      Chemin d'\'Ecogia, 1290 Versoix, Switzerland 
 \and Center for Astronomy and Space Sciences, University of 
      California, San Diego, La Jolla, CA 92093, USA
 \and Institut f\"ur Astronomie und Astrophysik, Universit\"at
      T\"ubingen, Sand 1, 72076 T\"ubingen, Germany
}

\date{Received  / Accepted}

\begin{document}

\abstract{We report on a series of outbursts of the high mass X-ray
  binary \src\ in 2010/2011 as observed with \textsl{INTEGRAL},
  \textsl{RXTE}, and \textit{Swift}. We discuss possible mechanisms
  resulting in the extraordinary outburst behavior of this source. The
  X-ray spectra can be described by standard phenomenological models,
  enhanced by an absorption feature of unknown origin at about 10\,keV
  and a narrow iron K$\alpha$ fluorescence line at 6.4\,keV, which are
  variable in flux and pulse phase. We find possible evidence for the
  presence of a cyclotron resonance scattering feature at about
  25\,keV at the 93\% level. The presence of a strong cyclotron line
  at 35\,keV seen in data from the source's 1998 outburst and
  confirmed by a reanalysis of these data can be excluded. This result
  indicates that the cyclotron line feature in \src\ is variable
  between individual outbursts.}

\maketitle
\section{Introduction}
Due to the $\sim$$10^{12}$\,G strong magnetic field at the magnetic
poles of many accreting neutron stars in high-mass X-ray binaries,
cyclotron resonance scattering features (CRSFs or cyclotron lines) are
observable in the X-ray spectra of these sources. These lines
originate from photons generated in the accretion column of a
magnetized neutron star interacting with electrons in the column,
since their motion perpendicular to the $B$-field is quantized into
Landau-levels with energy differences
\begin{equation}
\Delta E\approx 12\,\mathrm{keV}
\left(\frac{B}{10^{12}\,\mathrm{G}}\right) 
\end{equation}
To date CRSFs have been reported for about 20 X-ray pulsars
\citep{Caballero2011a}.

The $15.8\,\mathrm s$ pulsar \src\ was first detected in 1998
\citep{Smith1998a,Wilson1998a}. It is a transient X-ray source with a
Be-type companion \citep{Verrecchia2002a}. This kind of main-sequence
B stars shows Balmer emission lines caused by a quasi-Keplerian
equatorial disk near the Be star \citep[see,
e.g.,][]{Hanuschik1996a,Slettebak1988a}.  Interaction between this
disk and the orbiting neutron star can lead to violent X-ray outbursts
resulting in the appearance of a bright X-ray sources in the sky.

The initial outburst of \src\ in 1998 lasted about three months.
\citet{Heindl2001a} reported on the discovery of a CRSF at an energy
near $35\,\mathrm{keV}$ during this outburst, which was followed by
several fainter outbursts separated by $\sim$80\,d
\citep{Campana1999a}. This separation was later established by
\citet{Wilson2003a} as half of the $\sim$170\,d orbital period. The
occurrence of two outbursts per orbit could be related to the
misalignment of the Be star's angular momentum and the orbital plane
of the neutron star. While we look nearly onto the pole of the Be
star, the orbital inclination is $\gtrsim$$46^\circ$
\citep{Wilson2003a}.

\begin{figure}
 \includegraphics[width=\columnwidth]{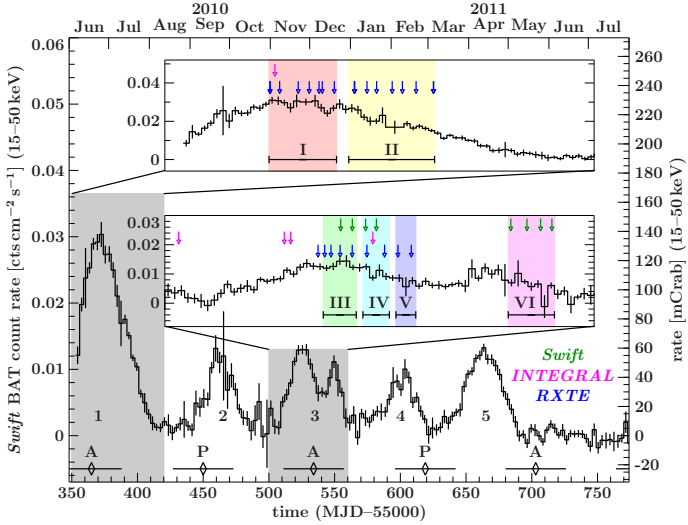}
 \caption{15--50\,keV \textit{Swift}/BAT light curve of the 2010/2011
   outburst series. The times of periastron and apastron passages are
   marked by P and A, respectively. These epochs and the corresponding
   uncertainties were calculated using the orbital solution from
   \citet{Wilson2003a}. The insets provide a closer view on the first
   (Jun/Jul) and the third (Nov/Dec) outburst. The blue, red, and
   green arrows in these insets indicate the observations times for
   PCA, ISGRI, and XRT, respectively. Epochs over which data were
   summed for the time resolved spectral analysis are indicated with~I
   to~VI.}
 \label{fig:lc}
\end{figure}

After a phase of nearly periodic flaring between 1998 and 2001, \src\
went into quiescence until 2010 June, when a new sequence of outbursts
started \citep[see, e.g.,][]{Mueller2010a}. The principal outburst
behavior in 2010 was the same as that observed in the 1990s, with two
outbursts per orbit. As shown in Fig.~\ref{fig:lc}, however, the five
outbursts seen during 2010 are neither clearly connected to the times
of periastron and apastron passages \citep[based on the orbital
ephemeris from][]{Wilson2003a}, nor to any other specific orbital
phase. This new outburst episode ended in 2011 June and the source was
back in quiescence at the time of writing.

In this paper, we report on \textit{INTEGRAL}, \textit{Swift}, and
\textit{RXTE} observations of the first and the third outburst of this
series in 2010 June/July and 2010 November/December, respectively. In
Sect.~\ref{sec:obs} we give a summary of the observations. In
Sect.~\ref{sec:timeres} we describe the continuum model and study the
time and pulse phase resolved behavior of the spectra. Summary and
conclusions are given in Sect.~\ref{sec:res}.

\section{Observations and data reduction}\label{sec:obs}

\begin{table}
\caption{Summary of all observations used.}\label{obs}

\begin{tabular}{lllrrr}
\hline\hline
ID\tablefootmark{a} & start date & MJD & $t_{\mathrm{exp}}$&
cts\tablefootmark{b} & e\tablefootmark{c}\\
 & (2010) & & [s] &[$10^{5}$] & \\
\hline
PCA & &\\
01-00 & Jun 20 & 55367.12--55367.14 & 1584 & 2.92 & I  \\
01-01 & Jun 20 & 55367.18--55367.20 & 1408 & 2.52 & I  \\
01-02 & Jun 21 & 55368.69--55368.77 & 4080 & 7.69 & I  \\
01-03 & Jun 24 & 55371.74--55371.78 & 2880 & 5.37 & I  \\
02-00 & Jun 26 & 55373.50--55373.61 & 5776 & 10.51& I  \\
02-03 & Jun 28 & 55375.14--55375.17 & 2800 & 5.04 & I  \\
02-01 & Jun 28 & 55375.66--55375.70 & 3184 & 5.62 & I  \\
02-02 & Jun 30 & 55377.62--55377.66 & 3200 & 5.36 & I  \\
03-00 & Jul 03 & 55380.89--55380.92 & 2000 & 3.06 & II \\
03-01 & Jul 03 & 55380.96--55380.99 & 3056 & 4.61 & II \\
03-02 & Jul 05 & 55382.92--55383.02 & 6368 & 8.96 & II \\
03-03 & Jul 07 & 55384.48--55384.59 & 5872 & 7.82 & II \\
04-00 & Jul 10 & 55387.03--55387.13 & 6144 & 7.46 & II \\
04-01 & Jul 11 & 55388.73--55388.77 & 3184 & 3.65 & II \\
04-02 & Jul 13 & 55390.95--55390.99 & 3200 & 3.37 & II \\
05-00 & Jul 16 & 55393.77--55393.80 & 3248 & 3.08 & II \\
05-01 & Jul 16 & 55393.83--55393.85 & 1392 & 1.28 & II \\
06-02 & Nov 23 & 55523.19--55523.23 & 2640 & 2.35 & ---\\
06-01 & Nov 24 & 55524.17--55524.24 & 3200 & 2.65 & III\\
06-00 & Nov 25 & 55525.15--55525.17 & 1392 & 1.21 & III\\
07-00 & Nov 26 & 55526.54--55526.56 & 2000 & 1.76 & III\\
07-01 & Nov 28 & 55528.36--55528.39 & 2656 & 2.35 & III\\
07-02 & Nov 30 & 55530.58--55530.62 & 2928 & 2.29 & IV \\
08-00 & Dec 03 & 55533.26--55533.30 & 3168 & 2.38 & IV \\
08-01 & Dec 05 & 55535.28--55535.31 & 2432 & 1.64 & V  \\
08-02 & Dec 07 & 55537.31--55537.34 & 2624 & 1.60 & V  \\
\hline
ISGRI & &\\
938 & Jun 20 & 55367.46--55368.44 & 45174 &   16.98 &  I \\
983 & Oct 31 & 55500.76--55503.45 & 35224 & $-$0.20 & ---\\
988 & Nov 18 & 55518.04--55518.16 & 4125  &    0.13 & ---\\ 
989 & Nov 18 & 55518.71--55519.42 & 10940 &    1.83 & ---\\
993 & Nov 30 & 55530.67--55532.35 & 54470 &    6.94 &  IV\\
\hline
XRT    & &\\
001 & Nov 26 &55526.60--55526.66 & 2475 & 0.13 & IV\\
002 & Nov 28 &55528.39--55528.47 & 2630 & 0.12 & IV\\
003 & Nov 30 &55530.39--55530.48 & 2424 & 0.12 & V \\
004 & Dec 02 &55532.00--55532.08 & 1442 & 0.07 & V \\
005 & Dec 22 &55552.49--55552.57 & 2600 & 0.10 & VI\\
006 & Dec 24 &55554.77--55554.98 & 1838 & 0.05 & VI\\
007 & Dec 26 &55556.90--55556.97 & 663  & 0.01 & VI\\
008 & Dec 28 &55558.31--55558.99 & 2110 & 0.05 & VI\\
\hline
\end{tabular}
\tablefoot{
  \tablefoottext{a}{For PCA, the first column contains the
    number of the Obs-ID after 95032-12- and for XRT after
    0031888. For ISGRI  the revolution number is listed.}
  \tablefoottext{b}{Total background corrected counts. Negative
    count rates are due to uncertainties in the background determination.}
  \tablefoottext{c}{Epoch for data grouping, see text for details. 
    ObsIDs without assignment in this column are different enough to
    other spectra during the same epoch that they cannot be combined
    with the other data or they were recorded without simultaneous
    low-energy measurements. We therefore exclude
    these data from our analysis.}}
\end{table}

The 2010/2011 outburst series started on 2010 June 4, when
\textit{Swift}/BAT detected an increase of the X-ray flux of \src,
rising up to 40\,mCrab (15--50\,keV) within three days
\citep{Krimm2010a}. This first outburst lasted about 60\,d and reached
a flux of $\sim$140\,mCrab (see Fig.~\ref{fig:lc}). The subsequent
four outbursts lasted between 30 and 50\,d, each, reaching almost the
same maximum flux level between 40 and 60\,mCrab. The separation of
the outbursts is between 60 and 90\,d. A peculiar behavior was
observed during the third outburst where the primary maximum was
followed by another brightening, reaching again luminosities up to
$\sim$50\,mCrab (Fig.~\ref{fig:lc}, inset).

We present data from the Proportional Counter Array
\citep[PCA,][]{Jahoda2006a} on board the \textit{Rossi X-ray Timing
  Explorer} \citep[\textit{RXTE},][]{Bradt1993a}, the
\textit{INTEGRAL} Soft Gamma-Ray Imager \citep[ISGRI,][]{Lebrun2003a}
on board the \textit{INTErnational Gamma-Ray Astrophysics Laboratory}
\citep[\textit{INTEGRAL},][]{Winkler2003a}, and the X-ray Telescope
\citep[XRT,][]{Burrows2005a} on board the \textit{Swift Gamma-Ray
  Burst Explorer} \citep{Gehrels2004a}. Data were reduced with the
standard analysis pipelines, based on \textsc{heasoft} (v.~6.10
and~6.11) and \textit{INTEGRAL} OSA v.~9.0. \src\ was monitored by
\textit{RXTE} regularly during the first and the third outburst. The
source was also sporadically in the field of view of \textit{INTEGRAL}
during the first and the third outburst. \textit{Swift} pointings were
available only during the main and secondary peak of the third
outburst. Table~\ref{obs} contains a log of the observations with
these satellites, which are also indicated in Fig.~\ref{fig:lc}.

PCA consisted of five proportional counter units (PCUs) with a field
of view of $\sim$$1^{\circ}$, sensitive between 2 and
$90\,\mathrm{keV}$. Since PCU2 is known to be the best calibrated one
\citep{Jahoda2006a}, only data from the top layer of this PCU are
used. We obtained 2--60\,keV light curves with 0.125\,s resolution,
spectra in the \texttt{standard2f} mode, and pulse phase resolved
spectra using \texttt{GoodXenon} data. The light curves were corrected
to the barycenter of the solar system using
\texttt{faxbary}\footnote{http://heasarc.gsfc.nasa.gov/lheasoft/ftools/fhelp/faxbary.txt}.
The PCA background model \texttt{SkyVLE} was used for PCA background
subtraction. Due to large uncertainties in the orbital parameters
\citep{Wilson2003a} no correction could be performed for the neutron
star's orbital motion. The lack of sufficient statistics prevented us
from improving the existing orbital solution. Data from the High
Energy X-ray Timing Experiment \citep[HEXTE,][]{Rothschild1998a} on
board \textsl{RXTE} were excluded from our analysis since both HEXTE
clusters were not rocking at the time of the observations. The
resulting uncertainties in the background determination are too large
for the purposes of this paper.

The CdTe detector of \textit{INTEGRAL}/ISGRI covers the energy range
from $\sim$18\,keV to 1\,MeV \citep{Lebrun2003a}. Thanks to the large
field of view of \textit{INTEGRAL}, \src\ was detected several times
during its recent outbursts between 2010~June and 2011~April in
observations pointed at Cyg X-1. We extracted ISGRI pulse phase
averaged spectra for all observations of \src\ using the standard
spectral extraction method of OSA~9 described in the \textit{INTEGRAL}
documentation\footnote{http://www.isdc.unige.ch/integral/analysis\#Documentation}.
We selected those observations of \src\ for which it was less than
$12^\circ$~off-axis.

\begin{figure}
 \includegraphics[width=\columnwidth]{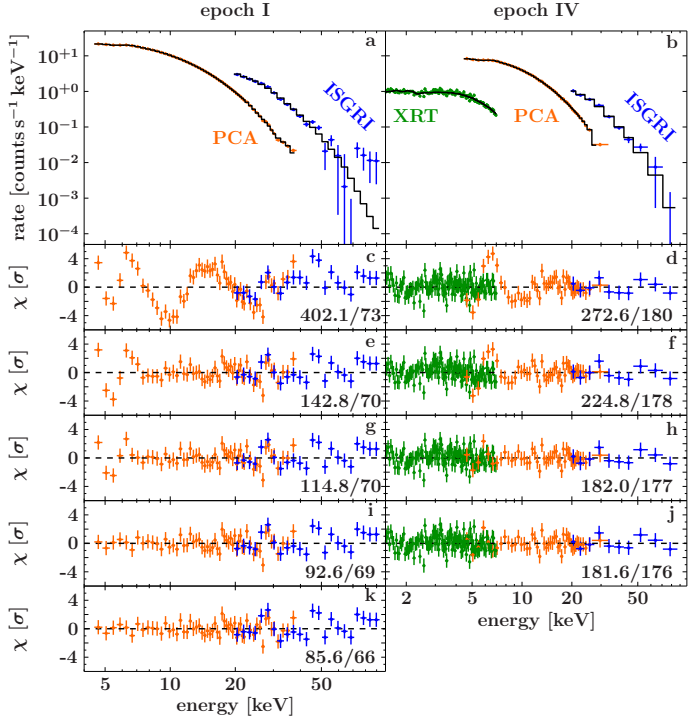}
 \caption{Panels~a and~b show two example spectra of \src.  Epoch~I is
   the spectrum with the highest countrate while epoch~IV is the only
   data set where all instruments are available.  Histograms show the
   best fit continuum. \mbox{Panels c--k} show the behavior of the
   residuals when adding the additional components of the spectral
   model one by one. Numbers at the bottom right of each panel
   indicate the best-fit statistics, $\chi^2/\mathrm{dof}$.  c and~d:
   best fit using the Fermi-Dirac cutoff only. e and~f: residuals
   after adding the 10\,keV feature to the continuum model. g and~h:
   residuals after adding the Galactic ridge emission to PCA (for
   epoch I the flux of this model component was fixed to the result
   obtained from epochs~III and~IV). i and~j: residuals after adding
   the Fe K$\alpha$ line to the model. Since in the model fit of epoch
   IV the depth of the CRSF results to zero, panel~j displays the
   final result for epoch~IV. In panel~k, the CRSF has been added to
   the data for epoch~I. See also text for a discussion of the
   statistical significances of the Galactic ridge emission, the
   10\,keV feature, and the CRSF.}
 \label{fig:spec}
\end{figure}

To cover the soft X-ray band we use data from \textit{Swift}/XRT, a
$600\times 600$ pixel CCD covering a field of view of $23\farcm6\times
23\farcm6$ in the energy range 0.2--10\,keV. We extracted the data in
Windowed Timing mode. For the source region we have choosen a circle
around the source with a radius of $\sim$$1'$. The background was
extracted using circles of the same radius at two off-source positions
about $2\farcm5$ away from the source.

In the spectral analysis, for XRT we use data between 1.5\,keV and
7.0\,keV. We binned the XRT and PCA data to a signal-to-noise ratio of
ten. Due to calibration problems of PCA at the Xe L-edge, we exclude
PCA data below 4.5\,keV from our analysis. Discarding bins with
$S/N<10$ results in an upper energy limit of typically 40\,keV for the
PCA. ISGRI covers the high energy band between 20\,keV and
100\,keV. We added a systematic error in quadrature to the PCA and the
ISGRI spectra using canonical values of 0.5\% and 2.0\%, respectively
\citep[see,][and IBIS Analysis User Manual]{Jahoda2006a}.

\section{Spectral analysis}
\label{sec:timeres}
We performed all fits using the \textit{Interactive Spectral
  Interpretation System} \citep[ISIS,][]{Houck2000a}. In order to
improve the signal to noise ratio of individual spectra we averaged
the data over six data blocks in time taking into account the flux
level and instrument availability. These epochs I--VI are defined in
Table~\ref{params} and shown in Fig.~\ref{fig:lc}. The spectra in
epochs~I and~II cover the first outburst at the maximum and fading
phase, respectively. Epochs III--V follow the maximum and the fading
phase of the third outburst. The flux in the maximum level of this
outburst (epoch~III) is comparable to the fading phase of the first
outburst (epoch~II). The last set of observations is summarized in
epoch~VI. These XRT data cover the fading phase of the flare right
after the third outburst. Before defining these epochs we confirmed
that the spectral variability during these epochs is negligible.

\begin{figure}
  \includegraphics[width=\columnwidth]{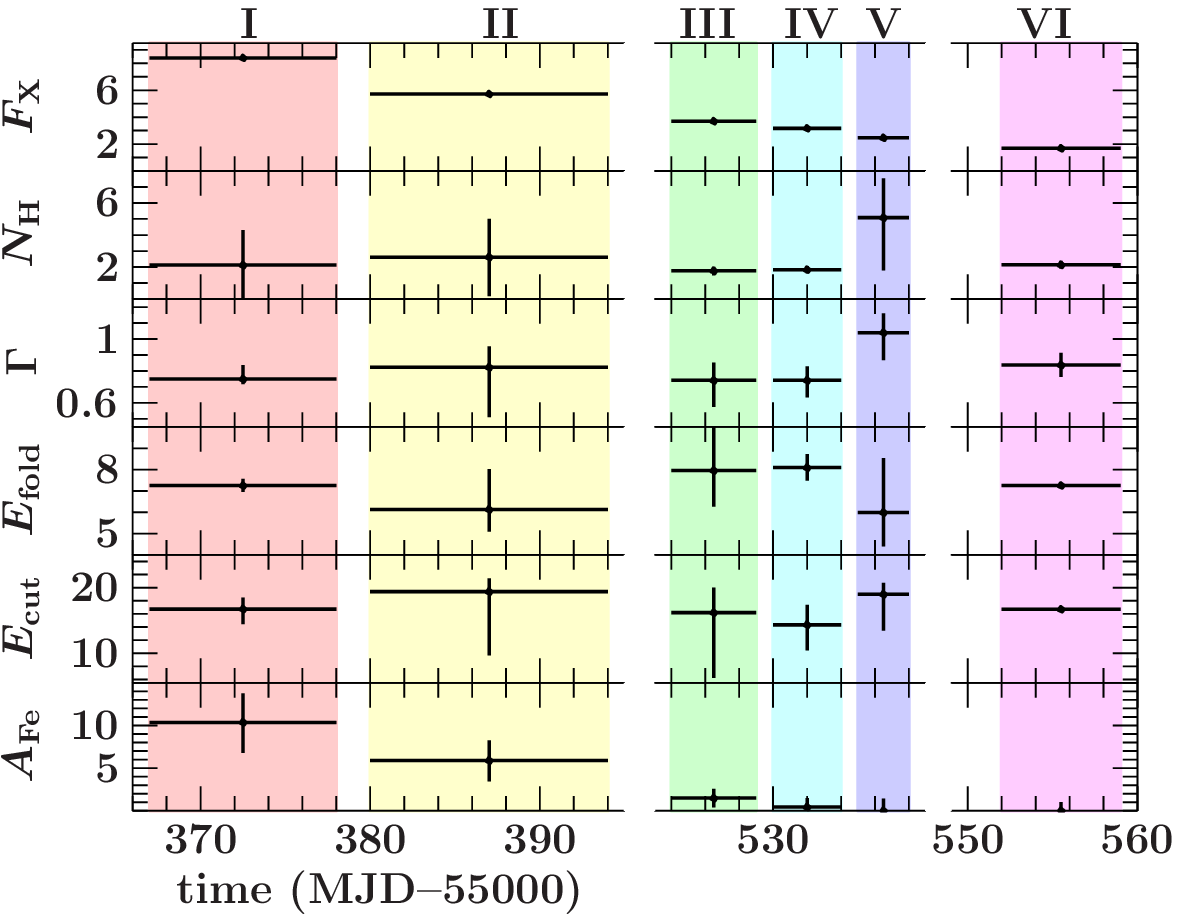}
  \caption{Results of time resolved spectroscopy. The units are as
    follows: $F_\mathrm{X}$ in $10^{-2}\,\LxE$ in the 7--15\,keV
    energy band, $N_\mathrm{H}$ in $10^{22}\,\mathrm{cm}^{-2}$,
    \Efold\ and \Ecut\ in keV, and $A_\mathrm{Fe}$ in
    $10^{-4}\,\LxE$.}
 \label{fig:TR}
\end{figure}

\begin{table*}
\caption{Results of the time resolved spectral analysis.}\label{params}
\begin{tabular}{l l l l l l l l l l l l l}
\hline\hline
e\tablefootmark{a} & start\tablefootmark{b}& stop\tablefootmark{b} & cts\tablefootmark{c} &
$F_1$\tablefootmark{d}&
$F_2$\tablefootmark{d}
&$N_\mathrm{H}$\tablefootmark{e} & $\Gamma$ & \Efold\tablefootmark{f}&
$E_\mathrm{cut}$\tablefootmark{f} &
$A_\mathrm{Fe}$\tablefootmark{g}&$W_\mathrm{Fe}$\tablefootmark{h}
&  \chisq/dof\\
\hline
I &367&378 &17.20 & $6.91(2)$&$8.40(2)$  & $2.1^{+2.2}_{-2.1}$  &$0.75^{+0.09}_{-0.04}$  &$7.25^{+0.31}_{-0.30}$
& $16.7^{+1.8}_{-2.4}$  & $10\pm 4$ & $55\pm 20$ &1.30/66 \\
II & 380&394 &16.46  &$4.71(3)$&$5.72(2)$ & $2.6^{+2.4}_{-2.5}$
& $0.82^{+0.14}_{-0.32}$  & $6.1^{+1.9}_{-1.0}$ &$19.4^{+2.1}_{-9.7}$
& $5.9^{+2.4}_{-2.5}$ & $46\pm20$ &1.03/45 \\
III & 524&529 &$\phantom{0}2.93$  & $3.03(3)$&$3.70(2)$  & $1.77^{+0.25}_{-0.29}$   &$0.74^{+0.12}_{-0.17}$
&$8.0^{+2.4}_{-1.7}$ &$16^{+4}_{-10}$  & $1.5\pm 1.0$&  $18\pm14$
  &1.17/202 \\
IV & 530&534 & $\phantom{0}1.68$  &$2.54(3)$&$3.17(2)$ & $1.84^{+0.22}_{-0.24}$
&$0.74^{+0.09}_{-0.10}$  &$8.1^{+0.7}_{-0.6}$ &$14\pm4$
&$0.4^{+1.0}_{-0.4}$ & $6^{+15}_{-6}$  &1.03/176\\
V & 535&538 &$\phantom{0}1.12$ &$1.88(2)$&$2.46(2)$ &$5.1^{+2.5}_{-3.3}$ &$1.04^{+0.13}_{-0.18}$
&$6.0^{+2.6}_{-1.6}$ &$19.0^{+1.8}_{-5.6}$ &$\leq 1.5$&$\leq 30$  &0.78/32 \\
VI & 552&559 &---& ---&$1.69(10)$ &$2.14^{+0.25}_{-0.24}$ &$0.84^{+0.08}_{-0.08}$ & 7.25
&16.7 &$\leq 1.0$&$\leq 27$ &1.21/135 \\
\hline\hline&$c_\mathrm{b}$&$c_{\mathrm{PCA}}$\tablefootmark{i}&$c_\mathrm{ISGRI}$\tablefootmark{i}&$c_\mathrm{XRT}$\tablefootmark{i}
&$E_\mathrm{G}$\tablefootmark{f} &
$\tau_\mathrm{G}$\tablefootmark{j} & $\sigma_\mathrm{G}$\tablefootmark{f} & $W_\text G$\tablefootmark{h}&
$E_\mathrm{CRSF}$\tablefootmark{f}
& $\tau_\mathrm{CRSF}$\tablefootmark{j} &$\sigma_\mathrm{CRSF}$\tablefootmark{f}&$W_\mathrm{CRSF}$\tablefootmark{h}\\
\hline
I &$0.943(20)$&1 &$0.97(3)$&--- & $9.85^{+0.20}_{-0.25}$  &
$6.9^{+2.7}_{-1.6}$  & $2.2^{+0.8}_{-0.5}$ & $-380^{+160}_{-80}$
&$25.3^{+0.9}_{-1.0}$&$9^{+10}_{-7}$&$0.65^{+1.46}_{-0.15}$& $-140^{+80}_{-120}$ \\ 
II&$1.00(7)$  &1 &---&--- & $9.9^{+0.4}_{-0.6}$  &
$3.9^{+4.8}_{-1.8}$  & $1.8^{+1.6}_{-1.0}$  & $-170^{+130}_{-200}$ &25.3 &$3^{+8}_{-3}$&0.65& $-60^{+60}_{-120}$ \\
III &$0.97(6)$ &1& ---&0.88(2)  &$9.8^{+0.9}_{-0.7}$  &
$3.9^{+2.0}_{-2.0}$  & 2 & $-180^{+100}_{-90}$  &25.3 & $\leq 23$& 
0.65 &  $\geq -330$ \\
IV&$0.97(4)$  &1 &$0.88(7)$ &$0.92(2)$  &$9.8^{+0.9}_{-0.9}$  &
$3.7^{+2.3}_{-2.3}$ & 2& $-170^{+100}_{-100}$ & 25.3 & $\leq 40$ &
0.65 &$\geq -500$ \\
V & 1  &1  &---&--- &---&---& --- &--- &--- &--- &--- &---   \\
VI&---  &  ---&---&1  & ---&---&---&--- &--- &--- &--- &---   \\
\hline
\end{tabular}
\tablefoot{
  Uncertainties and upper limits are at the 90\% confidence
  level. Numbers without error bars were held fixed at the
  values listed.\\
  \tablefoottext{a}{Epoch for data grouping.}
  \tablefoottext{b}{MJD$-$55000.}
  \tablefoottext{c}{Total background corrected PCA counts
    between 7 and 15\,keV, in multiples of $10^{5}$.}
  \tablefoottext{d}{Absorbed flux, in units of $10^{-2}\,\LxE$. $F_1$
    and $F_2$ cover the energy bands 10--20\,keV and 7--15\,keV,
    respectively. In
    epoch~VI, the 10--20\,keV band is not fully
    covered by the data and no flux value can be listed.}
  \tablefoottext{e}{In units of $10^{22}\,\mathrm{cm}^{-2}$.}
  \tablefoottext{f}{In units of keV.}
  \tablefoottext{g}{In units of
    $10^{-4}\,\mathrm{photons}\,\mathrm{s}^{-1}\,\mathrm{cm}^{-2}$. The
    centroid energy has been fixed to 6.4\,keV, and the width to
    $10^{-4}$\,keV.}
  \tablefoottext{h}{In units of eV.}
  \tablefoottext{i}{$c_\mathrm{PCA}$, $c_\mathrm{ISGRI}$ and
    $c_\mathrm{XRT}$ are defined as the cross calibration and
    normalization constants for PCA, ISGRI, and XRT, respectively.}
  \tablefoottext{j}{In multiples of $10^{-2}$.}
}
\end{table*}

\subsection{Spectral model}
As shown, e.g., by \citet{Becker2007a}, the X-ray spectra of
accretion-powered X-ray pulsars can be roughly described by a powerlaw
with a high energy cutoff. In practical data modelling, this continuum
has been approximated by several different continuum models
\citep[see, e.g.,][for a summary]{Kreykenbohm2002a}. Here we describe
the data using the so called Fermi-Dirac cutoff
\citep[FDCO,][]{Tanaka1986a}, given by
\begin{equation}
 \mathrm{FDCO}(E)\propto E^{-\Gamma}\times\left[1+\exp\left(\frac
     {E-E_{\mathrm{cut}}}{E_{\mathrm{fold}}}\right)\right]^{-1},
\end{equation}
which has been successfully applied to other accreting X-ray pulsars
such as Vela X-1 \citep{Kreykenbohm2008a}. This continuum is modified
by the CRSF, modeled as a line with a Gaussian optical depth profile
\begin{equation}\label{gauabs}
\exp(-\tau(E)),\ \mathrm{with} \
\tau(E)=\tau_\mathrm{CRSF}\times\exp\left[-\frac{1}{2}\left(\frac{E-E_\mathrm{CRSF}}{\sigma_\mathrm{CRSF}}\right)^2\right].
\end{equation}
The equivalent width of this feature will be denoted with
$W_\mathrm{CRSF}$ in the following. The spectra of some X-ray pulsars,
including \src, also contain an absorption or emission like feature in
the range 8--12\,keV \citep{Coburn2001a}. The origin of this so-called
10\,keV feature is still unclear, however, since it appears always at
about the same energy, it is probably not related to the magnetic
field strength of the neutron star. We modelled this feature as a
broad Gaussian absorption feature, as we also did for the CRSF
(Eq.~\ref{gauabs}) with the centroid energy, $E_\mathrm{G}$, the
width, $\sigma_\mathrm{G}$, optical depth, $\tau_\mathrm{G}$, and
equivalent width, $W_\mathrm{G}$. Finally, interstellar absorption was
modelled with an updated version of TBabs\footnote{see
  http://pulsar.sternwarte.uni-erlangen.de/wilms/research/tbabs/},
using abundances by \citet{Wilms2000a} and cross sections by
\citet{Verner1995a}.

In modelling the data, we have to take into account that the
cross-normalization of the different instruments used is not perfectly
well known and the source might also slightly change in flux between
the different observations. These effects were taken into account by
cross calibration constants $c_\mathrm{XRT}$ and $c_\mathrm{ISGRI}$,
using the PCA as the reference instrument. Furthermore, the PCA
background was allowed to vary slightly in count rate. To account for
this imperfections in the modelling of the background, we introduced
the constant $c_\mathrm{b}$.

Finally, the data modelling is affected by the proximity of the source
to the plane of the Galaxy. Galactic ridge emission \citep[GRE, see,
e.g.,][]{Worrall1982a,Warwick1985a} manifests itself through the
presence of an emission feature at \mbox{$\sim$6--7\,keV} in the PCA
spectrum caused by unresolved Fe K$\alpha$ fluorescence lines. These
lines are not present in the XRT data in epochs~III and~IV and
therefore must be due to diffuse emission that is picked up by the PCA
only, due to its larger field of view. In our modelling of the PCA
data we therefore introduced a model for the Galactic ridge emission
based on that of \citet{Ebisawa2007a}, who described the ridge
emission as the sum of two bremsstrahlung components and an iron line
complex modelled by three Gaussian lines at 6.4\,keV, 6.67\,keV, and
7.0\,keV, with equivalent width ratios of 85:458:129, respectively.
Since we used the PCA data down to 4.5\,keV only, we did not account
for the two soft bremsstrahlung components and modelled the Galactic
ridge emission as the sum of three narrow Gaussians, with fixed
energies and equivalent width ratios according to
\citet{Ebisawa2007a}. This component was applied only to the PCA data
and absorbed by the interstellar medium using the Galactic
$N_\mathrm{H}$ as determined from the Leiden/Argentine/Bonn (LAB)
Survey of Galactic H~\textsc{i}, \mbox{$N_\mathrm{H}=9.4\times
  10^{21}\,\mathrm{cm}^{-2}$} \citep{Kalberla2005a}. We determined the
flux of the Galactic ridge emission from simultaneous fits to the XRT
and PCA data from epochs III and IV. In epoch~IV, residuals around
6--7\,keV are only visible in PCA, while in XRT neither residuals from
this emission, nor from a source intrinsic iron K$\alpha$ line are
detected. Thus, the residuals in the epoch~IV PCA spectrum must be
caused by Galactic ridge emission. In epoch~III, on the other hand, we
find weak evidence for the presence of a source intrinsic iron
K$\alpha$ line in the XRT spectrum. The unabsorbed flux of the
6.4\,keV iron line of the Galactic ridge emission in both spectra is
consistent with each other (\mbox{$(7.2\pm
  2.0)\times10^{-5}\,\mathrm{photons}\,\mathrm{s}^{-1}\,\mathrm{cm}^{-2}$,}
and \mbox{$(6.6\pm
  2.0)\times10^{-5}\,\mathrm{photons}\,\mathrm{s}^{-1}\,\mathrm{cm}^{-2}$}
for epoch~III and IV, respectively).
While these values are slightly higher than the
$1.22\times10^{-5}\,\mathrm{photons}\,\mathrm{s}^{-1}\,\mathrm{cm}^{-2}$
reported by \citet{Ebisawa2007a}, the difference is still within the
typical variation of the Galactic ridge emission over the Galactic
plane \citep{Yamauchi2009a}. Since the flux of the Galactic ridge
emission is constant over time, we added this model component with
parameters fixed to the mean value as obtained from epochs III and IV
to the PCA spectra of all epochs. We find that for the late part of
the outburst, the ridge contributes 1.5\% of the \mbox{3--10keV} flux
and 8\% in the Fe band \mbox{(6--7keV).} To estimate the significance
of the Galactic ridge emission, we performed Monte Carlo simulations
of the best fit model without this feature to create a set of 1000
synthetic spectra. We then performed the fit allowing all model
parameters, including the Galactic ridge emission, to vary. For both
epochs~III and~IV none of these simulations led to a fake spectrum for
which the improvement in $\chi^2$ was as large as in the real data,
i.e., the probability that the Galactic ridge emission is real is
greater than 99.9\,\% ($>$$3.3\sigma$). The remaining residuals at
6.4\,keV can be explained by a narrow source intrinsic iron K$\alpha$
flourescence line. We model this feature by a thin Gaussian emission
line with fixed centroid energy $E_\mathrm{Fe}=6.4$\,keV and width
$\sigma_\mathrm{Fe}=10^{-4}$\,keV. The flux, $A_\mathrm{Fe}$, (and
thus equivalent width, $W_\mathrm{Fe}$) were allowed to vary.

In summary, the model used can be written as
\begin{equation} \label{eq:model}
M=\mathrm{TBabs}\times(\mathrm{FDCO}+\mathrm{Fe}_{6.4\,\mathrm{keV}})\times
G_{10\,\mathrm{keV}}\times G_{\mathrm{CRSF}}+\mathrm{GRE}.
\end{equation}

\subsection{Time resolved spectroscopy}\label{sect:trs}
In this Section we describe the time resolved behavior of the spectral
parameters. For each of the six epochs we fitted the respective
spectra separately, including the Galactic ridge emission as a
constant component as discussed above. Example spectra of two epochs
together with the best model fit are shown in Fig.~\ref{fig:spec}. The
free fit parameters are summarized in Table~\ref{params} and displayed
in Fig.~\ref{fig:TR}. We calculated fluxes in the energy band
\mbox{10--20\,keV} for the epochs \mbox{I--V}. In order to be able to
compare the source fluxes of all epochs, but avoid excessive
extrapolation of the models, we also derived the fluxes in the energy
band \mbox{7--15\,keV.} We excluded the contribution from the Galactic
ridge emission from this flux.

\begin{figure}
  \includegraphics[width=\columnwidth]{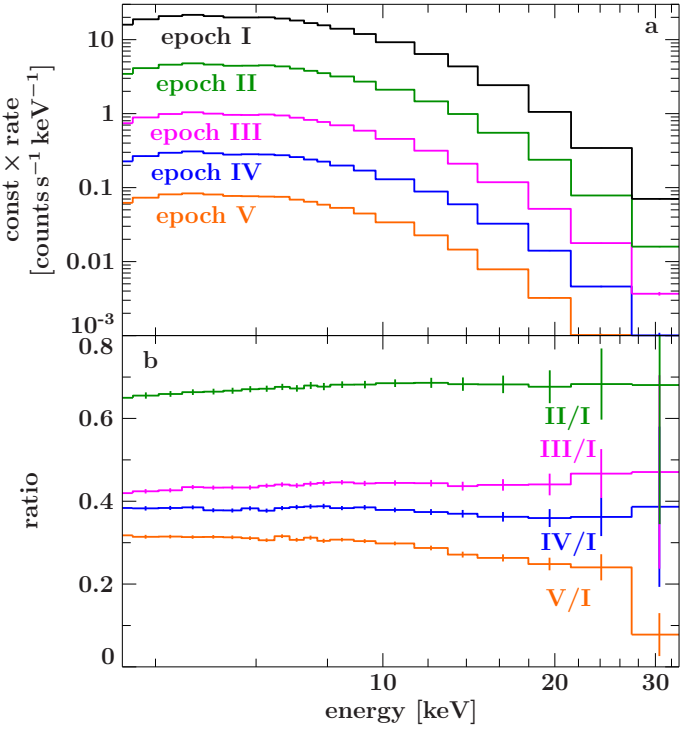}
  \caption{a: PCA spectra of time bins I--V (shifted in
    $y$-direction for better visibility). b: Ratio of the background
    corrected time resolved spectra.}
 \label{fig:TRratio}
\end{figure}

The resulting \chisq\ for all fits does not exceed 1.3 (see
Table~\ref{params}). For epoch~I, the value of \chisq\ is rather
high. Here, the greatest contribution to \chisq\ originates from
residuals of the ISGRI data which are caused by calibration problems,
so we accept this fit. Note that not all model components are
necessary to describe the data in the spectra with low statistics,
e.g., epoch~V provides an statistically too low \chisq\ of 0.78. The
reason for including these components in these fits as well, even
though the components overdetermine the fit model, is that this way it
can be shown that these spectra are at least consistent with the full
model. In addition, we note that an overestimation of the systematic
error would also yield a $\chi_\mathrm{red}$ that is too low. In
epoch~VI only XRT data are available and the continuum parameters are
badly constrained from these data alone. We therefore fixed \Efold\
and \Ecut\ to the value obtained from epoch~I because a change of
these parameters affects mainly energies not covered by XRT. However,
possible influences of these fixed parameters to the free fit
parameters $N_{\mathrm{H}}$ and $\Gamma$ might affect the results.

The behavior of the photon index, $\Gamma$, the cutoff energy, \Ecut,
and the folding energy, \Efold, yields information about the evolution
of the spectral continuum. In most cases there are no or only slight
variations of these parameters apparent. However, we know from
previous observations that these parameters can also be significantly
correlated to each other. We therefore also derived a model
independent illustration of the spectral changes by dividing the
background subtracted PCA spectra from \mbox{epochs II--V} by the
spectrum from epoch~I (see Fig.~\ref{fig:TRratio}). The ratios~II/I,
III/I, and IV/I appear to be mainly constants, meaning that the
variations of the continuum parameters in epochs \mbox{I--IV} are
probably artificial and due to cross correlations. In contrast to
this, ratio V/I shows a spectral softening for epoch~V, caused by a
real change of the continuum parameters.

\begin{figure}
  \includegraphics[width=\columnwidth]{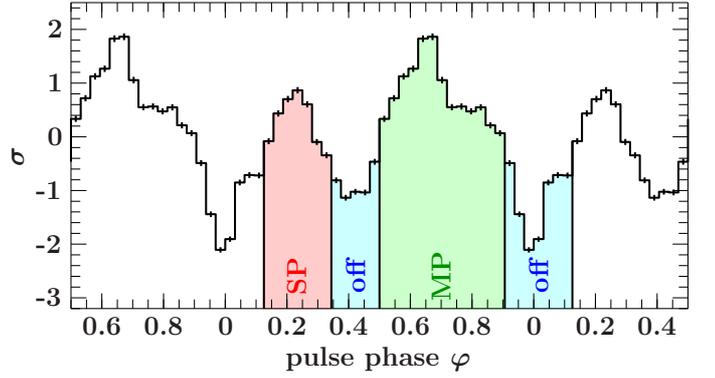}
  \caption{PCA-Pulse profile of epoch~I (in the full PCA energy band,
    i.e., 2--60\,keV), shown twice for clarity. The count rate in each
    bin is normalized to its variance relative to the mean pulse
    profile count rate. The phase bins are marked as follows. MP: main
    peak, SP: secondary peak, off: off state.}
 \label{fig:PPres}
\end{figure}

Except for epochs covered by XRT, the hydrogen column density
$N_\mathrm{H}$ is not well determined and the best-fit parameters are
consistent with constant $N_\mathrm{H}$. In addition to the line
caused by Galactic ridge emission, an Fe K$\alpha$ fluorescence line
at 6.4\,keV is required during the first outburst and the maximum
phase of the third outburst (epochs I--III). The flux of this line is
correlated with the X-ray flux of the source, as expected for a
fluorescent line. Furthermore, for these epochs at highest
luminosities (I--III), the equivalent width $W_\mathrm{Fe}$ stays
roughly constant. In the fading phase of the third outburst and its
subsequent flare (epochs IV--VI), this feature is consistent with
zero, meaning that the additional emission at these energies can be
explained by the Galactic ridge emission. To estimate the significance
of the Fe K$\alpha$ line, we performed similar Monte Carlo simulations
as those done for the Galactic ridge emission. For epochs~I and~II,
the probability that there is a source intrinsic Fe K$\alpha$ line is
greater than 99.9\,\% ($>$$3.3\sigma$). During epochs~III and~IV,
where the source was much fainter, these simulations show that the
probability for Fe line emission from the source is 98.5\%
($2.4\sigma$) and 62.0\% ($0.9\sigma$), i.e., here the Fe line region
is dominated by Galactic ridge emission.

We find no evidence for the centroid energy, the width, and the
optical depth of the 10\,keV feature to be variable over time.  Due to
the lack of statistics, the width of the feature, $\sigma_\mathrm{G}$,
in epoch~III and IV cannot be constrained and is therefore fixed to
the mean value obtained from epochs I and II, i.e.,
\mbox{$\sigma_\mathrm{G}=2.0$\,keV.} In epoch~V, also due to the lack
of statistics, this feature is not required to describe the data. We
therefore omit it from our model in this epoch, and also from
epoch~VI, where our coverage exists only below 7\,keV.

One of the most interesting questions is whether there is a cyclotron
line present as in the 1998 outburst \citep{Heindl2001a}. We find a
possible CRSF at $\sim$$25$\,keV during \mbox{epochs I--IV,} i.e.,
those epochs where good coverage exists above 10\,keV. The statistics
during the maximum of the first outburst (epoch~I) are good enough to
obtain CRSF parameters that are well constrained. As is fairly common
for CRSF fits \citep{Coburn2001a}, the width and depth of the line are
strongly correlated. We therefore set a lower limit of 0.5\,keV for
the width, which is comparable to the resolution of the PCA at these
energies. Monte Carlo simulations as described above lead to a 93\%
($1.81\sigma$) probability that the CRSF found in epoch~I is real.

\begin{figure}
  \includegraphics[width=\columnwidth]{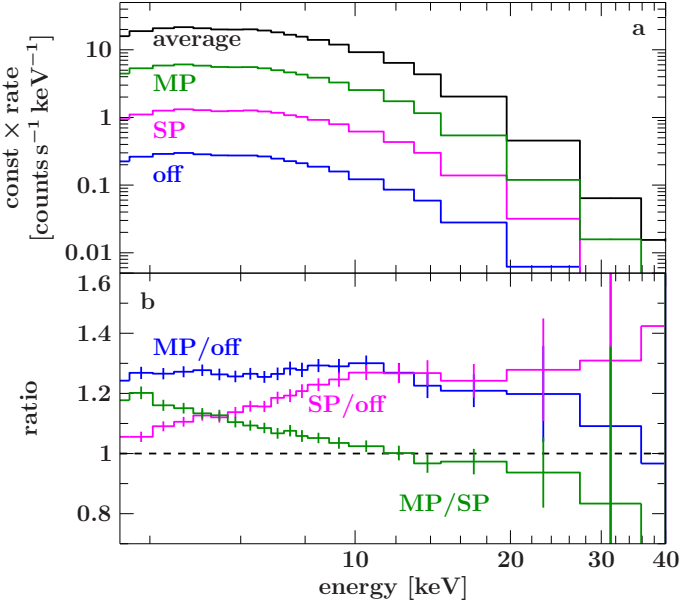}
  \caption{a: PCA spectra of the three phase bins together with the
    pulse phase averaged spectrum (shifted in $y$-direction for better
    visibility). b: Ratio of the background corrected pulse phase
    resolved spectra.}
 \label{fig:PPratio}
\end{figure}

The dependence of the CRSF's parameters to the choice of the approach
of modeling the continuum is an important point which also has to be
discussed. The results presented here were obtained by first adding
the 10\,keV feature to the model, and then accounting for the CRSF.
Doing this vice versa for epoch~I, we find another minimum of \chisq,
leading to different cyclotron line parameters, i.e.,
$E_\mathrm{CRSF}=29.6^{+1.6}_{-1.3}$\,keV,
$\tau_\mathrm{CRSF}=0.34^{+0.12}_{-0.09}$, and
$\sigma_\mathrm{CRSF}=6.8^{+1.2}_{-1.7}$\,keV. The main problem for
this alternative approach is the need of fixing parameters to certain
values for the final fit. For example, allowing $\sigma_\mathrm{CRSF}$
for the final fit to vary leads to an unrealistically broad CRSF which
effectively models part of the exponential rollover and not the line.
Furthermore, the quality of this fit is slightly worse ($\chi^2\approx
97$ vs.\ $\chi^2\approx 86$). We note that there is a third solution
in which the centroid energy of the CRSF is in agreement with
\citet{Heindl2001a}. However, contrary to \citet{Heindl2001a}, this
third solution has an unphysically broad and shallow shape and is thus
not physically meningful. All results presented in this paper are
based on the approach first adding the 10\,keV feature, and then the
CRSF. Finally, we note that there are two further solutions, where the
CRSF is located at 30\,keV and 40\,keV and a line width of less than
1\,keV. These two solutions are indeed physically meaningful, however,
a Monte Carlo estimation of the significances of these features yields
probabilities of 85\% ($1.44\sigma$) and 86\% ($1.48\sigma$),
respectively, much lower than for the solution with the CRSF at
25\,keV. Furthermore, these two solutions are supported only by about
three data bins per instrument (the 40\,keV solution even only by
ISGRI), while the 20\,keV solution is based on a much larger number of
spectral bins.

Including a CRSF in the fainter observations in epochs II--IV does not
significantly improve the fits. All observations are in principle
consistent with the presence of a weak line as that seen in
epoch~I. Including for consistency such a feature in the spectral
modeling, fixing the centroid energy and width and leaving the optical
depth respectively the equivalent width of the line as a free
parameter effectively gives an upper limit for the depth of the line
in these observations. Not unexpectedly, the limit for
$\tau_\mathrm{CRSF}$ becomes larger for the fainter phases of the
outburst (Table~\ref{params}). We find no evidence for the equivalent
width $W_\mathrm{CRSF}$ to be variable over time.

\begin{table}
\caption{Results of the pulse phase resolved spectral analysis.}\label{PPparams}

\begin{tabular}{l l l l l}
\hline\hline
parameter  & MP & SP  & off \\
\hline
$\varphi$ & 0.50--0.91 & 0.13--0.34  &0.34--0.50 \& 0.91--0.13 \\
$\Gamma$ &$0.87\pm 0.05$&$0.49^{+0.10}_{-0.13}$&$0.71^{+0.07}_{-0.09}$\\
\Efold\ [keV] &$6.7\pm 0.4$&$7.9^{+0.4}_{-0.5}$&$7.8\pm 0.5$\\
\Ecut\ [keV] &$19.2^{+1.2}_{-1.5}$&$11^{+4}_{-5}$&$15.0^{+2.4}_{-3.1}$\\
$A_{\mathrm{Fe}}\ [10^{-4}]$\tablefootmark{a} &$11.9\pm 2.5$ &
$8.7^{+2.8}_{-2.9}$ &$9.3\pm 2.2$ \\
$W_\mathrm{Fe}$ [eV] &$56^{+14}_{-12}$ &$46\pm15$ &$54\pm14$ \\
$E_{\mathrm{G}}$ [keV] &$9.7^{+0.6}_{-0.7}$&$9.5\pm0.3$&$9.9\pm 0.2$\\
$\tau_{\mathrm{G}}$
&$0.034^{+0.015}_{-0.014}$&$0.080^{+0.022}_{-0.020}$&$0.114\pm
0.018$\\
$W_\mathrm{G}$ [eV] &$-200^{+40}_{-20}$ &$-170^{+60}_{-20}$ & $-600^{+90}_{-80}$ \\
$\tau_{\mathrm{CRSF}}$&$0.12\pm 0.08$&$0.04^{+0.09}_{-0.04}$&$0.13\pm
0.09$\\
$W_\mathrm{CRSF}$ [eV] & $-180\pm 90$&$-60^{+60}_{-140}$ &$-210\pm 130$ \\
\hline
$\chi^2_{\mathrm{red}}$/dof&0.92/56 & 1.10/55 &0.93/56 \\
\hline
\end{tabular}
\tablefoot{
  Uncertainties and upper limits correspond to the 90\% confidence
  level for one parameter of interest.\\
  \tablefoottext{a}{In units of
    $\mathrm{photons}\,\mathrm{s}^{-1}\,\mathrm{cm}^{-2}$. The
    centroid energy has been fixed to 6.4\,keV, and the width to
    $10^{-4}$\,keV.} 
}
\end{table}

\subsection{Pulse phase resolved spectroscopy}
\label{sec:phaseres}

Spectral parameters are known to be variable as a function of pulse
phase, e.g., the parameters of cyclotron lines \citep[e.g., 1A
1118$-$61;][]{Suchy2011a}. To investigate such changes, we performed
pulse phase resolved spectroscopy on \mbox{epoch~I} with PCA, based on
a pulse period ephemeris found by epoch folding. Epoch I provides the
best statistics for this analysis. The ephemeris can be described with
a constant spin up, \mbox{$\dot P_\mathrm{pulse}=(-3.0\pm 0.3)\times
  10^{-9}\,\mathrm{s}\,\mathrm{s}^{-1}$} with pulse period between
15.755\,s and 15.767\,s. The uncertainties of the pulse periods were
typically less than 0.003\,s. Since it was not possible to correct the
light curve for the orbital motion of the system, this spin up trend
might be caused by orbital effects and not by transfer of angular
momentum of the accreted matter onto the neutron star. We extracted
pulse phase resolved spectra using individual pulse periods following
this ephemeris. The resulting pulse profile (Fig.~\ref{fig:PPres})
shows a double peaked structure, typical for such sources \citep[see,
e.g.,][]{Bildsten1997a}.

As in the time resolved case, we used the FDCO model to fit the phase
resolved spectra using three phase bins as indicated in
Fig.~\ref{fig:PPres}: The main peak (MP), the secondary peak (SP), and
the off state (off). The pulse phases of the bins used are summarized
in Table~\ref{PPparams}. Constant Galactic ridge emission is taken
into account in the same way as for the pulse averaged analysis. The
background scaling factor for PCA $c_\mathrm{b}$ is assumed to be
equal to the result from the pulse averaged analysis and was therefore
frozen to this value. Initial fits of the pulse phase resolved spectra
lead to large unconstraints in $N_\mathrm{H}$ and the width of the
10\,keV feature $\sigma_\mathrm{G}$. These values were therefore
frozen to the results form the phase averaged analysis. The signal to
noise ratio in the spectra from the individual pulse phases is too low
to allow us to study the behavior of the CRSF. Similar to the flux
dependent analysis, we freeze the CRSF centroid energy and width to
the result from the pulse averaged analysis and allow only the optical
depth respectively the equivalent width to vary (due to the marginal
significance of the CRSF no further attempt was made to study a phase
dependence of the cyclotron line).

\begin{figure}
  \includegraphics[width=\columnwidth]{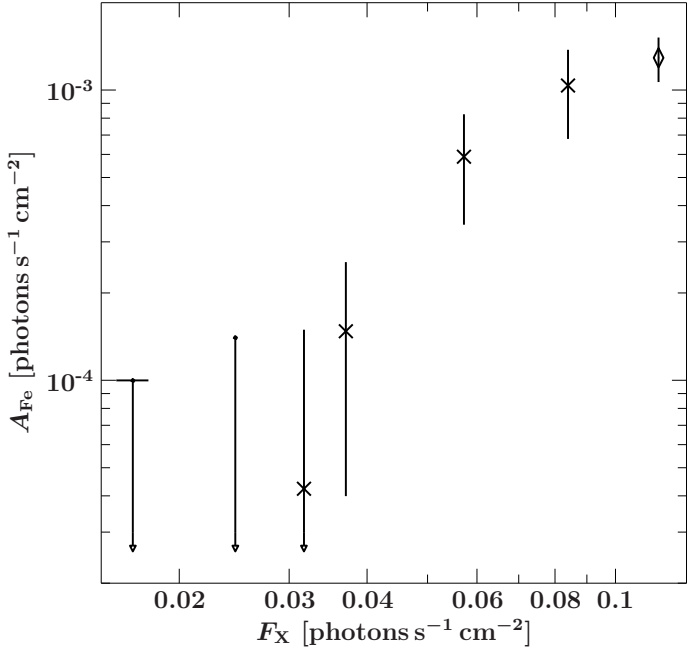}
  \caption{Flux of the iron K$\alpha$ fluorescence line against the
    7$-$15\,keV flux. The crosses correspond to the data from the 2010
    outburst, and the diamond to the average spectrum of the 1998
    outburst, obtained by reextracting and analyzing the old PCA and
    HEXTE data from 1998.}
 \label{fig:Fe}
\end{figure}

Figure~\ref{fig:PPratio} shows the ratios of background corrected
pulse phase resolved spectra. The ratio MP/off appears mainly as a
constant, while the other two ratios reveal more complex shapes. This
constant ratio is in agreement with the spectral parameters of the
main peak and the off state. These parameters show almost no or only
slight differences to each other. The ratio MP/SP, a falling line, can
be explained by a change in photon index. This explanation is
confirmed from the spectral analysis. While \Ecut\ and \Efold\ show
only small differences, the photon index $\Gamma$ is significantly
lower in the secondary peak, i.e., the spectrum in this phase bin is
harder than in the other phase bins. The centroid energy of the
10\,keV feature does not vary with pulse phase while its depth weakens
during the main peak. The equivalent width remains constant in the
main and the secondary peak, while it shows a larger value in the off
state. Finally, we note that there is neither evidence for changes of
the Fe line flux nor for the equivalent width over the pulse phase.

\section{Results and conclusions}
\label{sec:res}

\subsection{Outburst series}
In this paper we analyzed quasi-simultaneous XRT, ISGRI, and PCA
observations of two outbursts of \src\ during a series in the second
half of 2010.

\begin{figure}
  \includegraphics[width=\columnwidth]{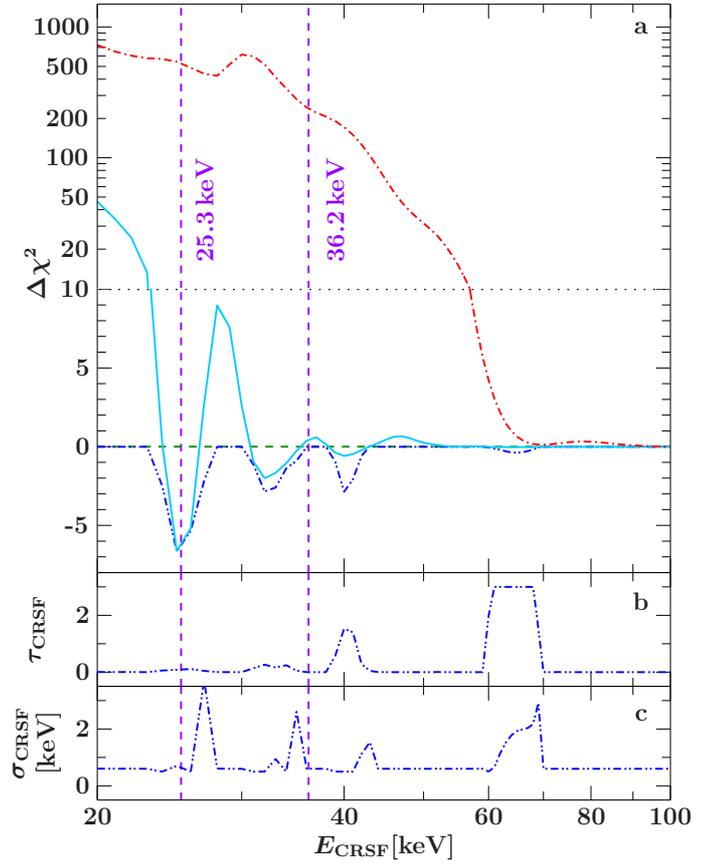}
  \caption{a: Difference of $\chi^2$ of the best fit of epoch~I with
    frozen CRSF parameters (Fig. \ref{fig:spec}k) and the $\chi^2$ of
    the best fit without CRSF. The centroid energy has been varied in
    steps of 1\,keV. The red, dashed-dotted curve displays the
    $\chi^2$ of the best fit when using $\tau_\mathrm{CRSF}$ and
    $\sigma_\mathrm{CRSF}$ from \citet{Heindl2001a}. For the light
    blue, solid curve the parameters from this work have been
    used. The dark blue, dashed-doubledotted curve shows this
    difference when both, the width and depth (the upper limit has
    been set to 3) of the CRSF are allowed to vary, i.e., the same
    situation as that of Fig.~\ref{fig:spec}. The two additional
    minima in $\chi^2$ are discussed in Sect.~\ref{sect:trs}. The
    horizontal dashed line indicates $\Delta\chi^2=0$, and the
    horizontal dotted line seperates the logarithmic and the linear
    scaling in $y$-direction.  The vertical dashed lines shows the
    position of the CRSF derived in \citet{Heindl2001a} and this
    work. \mbox{b and c:} results for the optical depth and the width
    of the CRSF, respectively, when allowing these parameters to
    vary. The plateau of the curve displayed in panel b at
    \mbox{$\sim$60--70\,keV} is caused by the upper limit for
    $\tau_\text{CRSF}$ of 3.}
 \label{fig:chisq}
\end{figure}

Before the onset of this outburst series, \src\ was in a state of
quiescence for almost one decade, i.e., about 20 orbits of the neutron
star. Even though the formation, structure, and dynamics of Be disks
are even today far from being completely understood \citep[see,
e.g.,][and references therein]{Draper2011a}, the missing mass
accretion onto the neutron star during that time is probably due to
the absence of a sufficiently large Be disk during that time.  A new
outburst was only possible once the disk had been replenished and
accretion could be triggered.

This series shows similar behavior as the one observed in 1998, e.g.,
two outbursts are observed per orbital period. The outbursts of \src\
could also be similar to those seen in GX~301$-$2, which have been
extensively modeled by \citet{Leahy2002a}. These authors posit that an
additionally stream of matter is flowing from the primary, and that a
second outburst per orbit could be caused by the passage of the
neutron star through this stream. These flux peaks of GX~301$-$2 occur
near the apsides of this system.

However, the outbursts of \src\ do not clearly coincide with the times
of periastron and apastron passages of the neutron star
(Fig.~\ref{fig:lc}). This could be explained by 3-dimensional
simulations, which show that the disturbance of the Be disk by the
gravitational field of the neutron star could lead to a strong
asymmetric structure of the circumstellar material, which could also
lead to multiple X-ray outbursts during one orbital period \citep[see,
e.g.,][]{Okazaki2011a}.

Alternatively, the outbursts can be triggered by the neutron star
passing through the Be disk due to a misalignment of the orbit and the
Be star's equatorial plane \citep{Wilson2003a}.

Another possible explanation for the irregularity of the outbursts is
that they are triggered by density variations in the Be disk and not
by orbital effects alone. This assumption could be verified by regular
optical monitoring of the Balmer $\text H\alpha$ line, which is an
indicator for the presence of such a disk \citep[see, e.g.,][and
references therein]{Grundstrom2007a}.

\subsection{Spectroscopic results}
An absorbed Fermi-Dirac cutoff powerlaw together with an iron $\text
K\alpha$ fluorescence line, an iron line complex between 6 and 7\,keV
caused by Galactic ridge emission, a Gaussian like absorption feature
around 10\,keV, and a cyclotron line at $\sim$25\,keV reproduce the
observations well in terms of \chisq. We find time as well as pulse
phase dependent variability of the continuum parameters of \src. In
the time resolved case, these changes might be caused by different
accreting mechanisms depending on the mass transfer onto the neutron
star. The periodically changing line of sight with respect to the
X-ray emitting region, caused by the rotation of the neutron star, is
likely responsible for the variabilities observed in the pulse phase
resolved analysis. Furthermore, this rotation together with
differences in the accretion geometries at the two magnetic poles of
the neutron star might lead to the asymmetric pulse profile of the
X-ray pulsar (see Fig.~\ref{fig:PPres}).

Studying the behavior of the spectral shape is problematic, because
the spectral parameters photon index $\Gamma$, folding energy \Efold,
as well as the cutoff energy \Ecut, which describe the broadband shape
of the X-ray spectrum, show strong cross correlations to each other.
To account for this, we calculated ratios of the spectra for both, the
time resolved and the pulse phase resolved studies. In the time
resolved case, we find clear deviations from a constant ratio for
epoch~V. During epochs I--IV, the overall spectral shape seems to
remain relatively constant. In the pulse phase resolved analysis, the
spectrum of the secondary peak turns out to be significant harder than
these from the main peak and the off state. Such properties of X-ray
spectra are due to, e.g., the temperature of the visible part of the
X-ray emitting region, in particular the accretion column.

Another feature required to get a good fit is a source intrinsic
Gaussian iron K$\alpha$ emission line at 6.4\,keV. The disagreement in
the strength of the required emission in this energy range between PCA
and XRT in initial fits can be solved by inducing the emission of the
galactic ridge. The strength of the Galactic ridge emission is about
four times larger than found in other regions of the Galaxy
\citep{Ebisawa2007a}. Such a difference is consistent, however, with
the typical spatial variations of the ridge emission
\citep{Yamauchi2009a}. Furthermore, \citet{Kuehnel2012a} found an
emission strength consistent with our result for GRO~J1008$-$57.

After taking into account the contribution due to Galactic ridge
emission, the flux of the source intrinsic Fe K$\alpha$ emission line
is correlated with the X-ray flux $F_\mathrm{X}$ (see
Fig.~\ref{fig:Fe}), as also observed in other X-ray transients
\citep[see, e.g.,][]{Inoue1985a}. The Fe K$\alpha$ line is
significantly detected at source fluxes \mbox{(7--15\,keV)} greater
than about 0.035\,\LxE. At fluxes below this value, the source
intrinsic line is consistent with zero. The equivalent width
$W_\mathrm{Fe}$ stays constant for high luminosities, i.e., epochs
I--III. Reanalyzing the earlier \textit{RXTE} data \citep{Heindl2001a}
with the continuum model employed here yields an equivalent width
which is also in agreement with the result from the 1998 outburst
($W_\mathrm{Fe}^{1998}=59\pm10$\,eV). The different values for
$W_\mathrm{Fe}$ for the fainter observations, where the Fe K$\alpha$
line is only marginally detected, can be caused by the uncertainties
of the Galactic ridge emission, which significantly contributes to the
data at these energies. In the pulse phase resolved analysis we find
no evidence for variability of the line flux. This result indicates
that it is emitted in a region large compared to the distance
travelled by light during one pulse period. The equivalent width also
shows no variation, which we would actually expect for a constant
model component at varying flux levels. However, this can be explained
by the quite large relative uncertainty of $W_\mathrm{Fe}$, which is
on the same order of magnitude as the respective variations.

The 10\,keV feature is present during all epochs and pulse phases. It
shows relatively constant results for line energy, width, and depth in
the time resolved case. The reanalysis of the 1998 data yields an
equivalent width of $W_\mathrm{G}^{1998}=-250^{+120}_{-270}$\,eV,
which also equals the results for the current outburst. While we find
almost no connection of the parameters of this feature with the
luminosity, its optical depth and equivalent width varies with pulse
phase. $W_\mathrm{G}$ is constant during the main and the secondary
peak, but it strongly increases during the off state. This behavior
indicates that the fractional amount of absorbed flux related to the
continuum level is constant for the main and the secondary peak and
changes in the off state. This behavior of the 10\,keV feature could
give rise to speculations about possible physical processes as, e.g.,
pulse phase dependent absorbing processes, producing such a
feature. On the other hand, the residuals around 10\,keV could also be
due to the failing of the spectral broadband continuum models
resulting in a wrong description of the data around these energies.
More quantitative analyses of this feature, also from other sources
where it occours in emission, are urgently needed to reveal the true
nature of this enigmatic feature.

\subsection{Cyclotron resonance scattering feature}

We find weak evidence for the presence of the CRSF, first discovered
by \citet{Heindl2001a} during the 1998 outburst series.  The cyclotron
line improved the model fit for the high signal to noise epoch~I
spectrum, where PCA and ISGRI data are available.  Later spectra are
consistent with the presence of a CRSF with unchanged parameters,
however, the line is not formally detected in these observations
because of their lower signal to noise ratio. We stress that due to
its relative weakness, the inclusion of the CRSF in our fits does not
affect our results for the continuum parameters or for the pulse phase
resolved analysis.

If the identification of the line is correct, then its depth in 2010
was significantly lower than that measured in the 1998 outburst
\citep[$\tau_\mathrm{CRSF}^{2010}=0.09^{+0.10}_{-0.07}$ vs.\
$\tau_\mathrm{CRSF}^{1998}=0.33^{+0.07}_{-0.06}$,][]{Heindl2001a}.
Furthermore the centroid energy of the CRSF in 2010
($E_\mathrm{CRSF}^{2010}= 25.3^{+0.9}_{-1.0}$\,keV) is significantly
lower than in 1998 \citep[$E_\mathrm{CRSF}^{1998}=
36.2^{+0.5}_{-0.7}$\,keV,][]{Heindl2001a}. Reanalyzing the earlier
\textsl{RXTE} data as described above yields CRSF parameters which are
consistent with those found by \citet{Heindl2001a}.  Furthermore, the
equivalent width in 1998 was also significantly larger than in the
current data \mbox{($W_\mathrm{CRSF}^{1998}=(-2.1^{+0.6}_{-1.0})\times
  10^{3}$\,eV).}

One reason why constraining the CRSF is so much easier in the 1998
data is due to the fact that, while the exposure times of the 1998 and
the 2010 data are comparable, the 10--20\,keV X-ray flux of the 1998
observation is $\sim$40\% higher than in the 2010 observation
(\mbox{$\sim$$9.6\times
  10^{-2}\,\mathrm{photons}\,\mathrm{s}^{-1}\,\mathrm{cm}^{-2}$} vs.\
\mbox{$\sim$$6.9\times
  10^{-2}\,\mathrm{photons}\,\mathrm{s}^{-1}\,\mathrm{cm}^{-2}$}).
However, fits to the 2010 data in which we fix the CRSF parameters to
their 1998 values do not result in a satisfactory description of the
data ($\chi^2_\mathrm{red}=1.55$). Could this indicate that the energy
of the line varied? Assuming that the CRSF width and depth remained
the same, we searched for a line similar to that found in the 1998
data. Varying the CRSF energy in steps of a 1\,keV and refitting the
continuum parameters did not yield any satisfactory results, with
best-fit $\chi^2_\mathrm{red}>1.5$ over the 20 to 60\,keV band (see
Fig. \ref{fig:chisq}, red curve). This result indicates that
irrespective of the difference in signal to noise between both data
sets the CRSF must have varied between both outbursts, and a strong
line as that seen in 1998 is not consistent with the data analyzed
here.

Taking the 2010 CRSF values at face value, a possible explanation for
the difference between the 1998 and 2010 outbursts could be the flux
dependence that is seen in some other CRSF sources \citep[see,
e.g.,][for a recent discussion]{Caballero2011a}. \src\ could therefore
be a cyclotron source with an overall positive correlation between the
X-ray flux and the CRSF energy, similar to Her X-1
\citep{Staubert2007a}. In models for the change in cyclotron line
energy in neutron stars this change is generally interpreted as a
change in height of the line producing region as the accretion column
changes with mass accretion rate (and thus luminosity). Using the
dipole approximation, and the assumption that in 1998 the CRSF was
emitted from the neutron star's surface ($r_\mathrm{NS}=10$\,km and
$m_\mathrm{NS}= 1.4\,\text M_\odot$) yields a height difference of
1.2\,km for the regions where the CRSF is generated. This result is in
agreement with typical estimates of several kilometers for the height
of the accretion columns \citep[see, e.g.,][]{Basko1976a}.
\citet{Becker2012a} show that a positive correlation between flux and
energy is possible in the luminosity range where the stopping in the
accretion column is dominated by Coulomb braking. For cyclotron line
energies around 30\,keV Coulomb braking is the dominating braking
process in the luminosity range
\mbox{1--$5\times10^{37}\,\mathrm{erg}\,\mathrm{s}^{-1}$.} For higher
luminosities, radiation braking dominates. \src's peak outburst
luminosity of $4.5\times 10^{37}\,\mathrm{erg}\,\mathrm{s}^{-1}$ is
barely consistent with this range, however, given that the source
seems to be located at the transition between Coulomb and radiation
braking and was brighter in 1998, the very large change in CRSF energy
seems unlikely to be due to a pure mass accretion rate effect.

In conclusion, while the possibility of a luminosity dependent CRSF is
intriguing, the poor signal to noise ratio of the 2010 data does not
allow a definitive answer concerning the luminosity dependence of the
CRSF. Further, longer, monitoring observations of \src\ during its
next outburst episode are urgently needed to resolve this question.

\begin{acknowledgements}
  The authors thank J\'er\^ome Rodriguez for his help with the
  \textit{INTEGRAL} observations, and the schedulers of \rxte\ and
  \textit{Swift} for their role in making this campaign possible. We
  thank the anonymous referee for his/her thorough review of this
  paper and his/her constructive comments.  We thank John E. Davis for
  the development of the SLxfig module, which was used to create all
  figures in the paper. We acknowledge funding by the
  Bundesministerium f\"ur Wirtschaft und Technologie under Deutsches
  Zentrum f\"ur Luft- und Raumfahrt grants 50~OR~0808, 50~OR~0905, and
  50~OR~1113. I.C. acknowledges financial support from the French
  Space Agency CNES through CNRS. This research is also based on
  observations with \textit{INTEGRAL}, an ESA project with instruments
  and science data centre funded by ESA member states (especially the
  PI countries: Denmark, France, Germany, Italy, Switzerland, Spain),
  Czech Republic, and Poland, and with the participation of Russia and
  the USA.
\end{acknowledgements}

\bibliography{mnemonic,aa_abbrv,references}
\bibliographystyle{aa}

\end{document}